\documentclass[12pt]{article}
\usepackage[top=1.15in, bottom=1.4in, left=1in, right=1in]{geometry}
\usepackage{booktabs}

\usepackage{cite}
\usepackage{enumerate,setspace,graphicx,epstopdf,amsmath,amsfonts,amssymb,amsthm}
\usepackage{marginnote,datetime,enumitem,subfigure,rotating,fancyvrb}
\usepackage[hang]{footmisc} 
\usepackage{authblk}
\usepackage{threeparttable}
\usepackage{natbib}
\usepackage{indentfirst}
\usepackage{epsfig}
\usepackage{graphicx}
\usepackage{subcaption}
\usepackage{rotating}
\usepackage{float}

\usepackage{cite}
\makeatletter
\renewcommand\@cite[1]{\textsuperscript{#1}}
\makeatother

\begin{document}

\title{\vspace{-2.5cm} \textbf{Social media and suicide: empirical evidence from the quasi-exogenous geographical adoption of Twitter}}

\author[1]{Alexis Du}
\author[2]{Thomas Renault}
\affil[1]{Universit\'e Paris 1 Panth\'eon-Sorbonne}
\date{\today}

\onehalfspacing
\maketitle
\vspace{-0.25in}

\begin{abstract}

Social media usage is often cited as a potential driver behind the rising suicide rates. However, distinguishing the causal effect—whether social media increases the risk of suicide—from reverse causality, where individuals already at higher risk of suicide are more likely to use social media, remains a significant challenge. In this paper, we use an instrumental variable approach to study the quasi-exogenous geographical adoption of Twitter and its causal relationship with suicide rates. Our analysis first demonstrates that Twitter’s geographical adoption was driven by the presence of certain users at the 2007 SXSW festival, which led to long-term disparities in adoption rates across counties in the United States. Then, using a two-stage least squares (2SLS) regression and controlling for a wide range of geographic, socioeconomic and demographic factors, we find no significant relationship between Twitter adoption and suicide rates. 

\vspace{3mm}

\end{abstract}

\thispagestyle{empty}

\clearpage

\doublespacing

\section{Introduction}
\label{intro}

Between 2001 and 2021, the age-adjusted suicide rate in the United States increased from 10.7 to 14.1 deaths per 100,000 standard population, marking a 31.8\% rise over the two decades. This period also coincides with the rapid development of the Internet and social media, and several surveys or cross-sectional studies indicate that frequent social media users tend to experience issues with anxiety, depression, and sleep disorders \citep{ortiz2019facebook, twenge2019media, lee2022social}. Social media may contribute to higher suicide rates through various mechanisms \citep{luxton2012social}, such as cyberbullying, the facilitation of suicide pacts, increased exposure to suicide-related news, and the amplification of extreme or harmful opinions. However, it is important to recognize that social media can also have positive effects by providing access to support networks, facilitating mental health interventions, and connecting individuals to resources that promote well-being. In the end, the causal effect remains uncertain, underscoring the importance of measuring and quantifying these impacts carefully.

One of the major difficulties in the literature on the impact of social media is the risk of reverse causality, as it is challenging to ascertain whether social media usage is the cause or the effect of certain outcomes \citep{sabatini2023behavioral}. For example, individuals who already experience anxiety or depression may turn to social media as a coping mechanism or a way to seek support, which could inflate the observed association between social media use and mental health issues. Conversely, heavy social media use might contribute to the worsening of mental health by exposing users to cyberbullying, social comparison, or sleep disturbances. 

To identify the causal effect of social media using real platform data, researchers must find a (quasi-)exogenous source of variation in either the availability or demand for social media. In terms of availability, this could include bans (such as the recent ban of X in Brazil), age restrictions (like China’s policy on limiting Tiktok usage to 60-minutes daily for those who are under 18), or staggered introduction (as for the development of Facebook across U.S. colleges). For demand-side variations, one promising identification strategy is based on the centrality of early adopters (see \cite{enikolopov2020social} on the geographical differences in early adopters of the VK social media platform in Russia and their impact on protests). This approach relies on the peer effect in product adoption \citep{bailey2022peer} as social influence can lead to clustering of product adoption within certain groups or geographic areas, as people are more likely to adopt a product when they see others around them using it.  

In this paper, we propose to leverage the quasi-exogenous geographic expansion of Twitter, driven by the presence of users from various U.S. counties at the South by Southwest (SXSW) festival in 2007. This instrument was used by \cite{muller2023hashtag} to analyze the link between social media and hate crimes, and by \cite{fujiwara2024effect} to study the relationship between social media and voting in elections. We first confirm that this event lead to long-term variations in the geographical adoption of Twitter in the United States. Twitter adoption grew at a significantly higher rate in counties where many new SXSW followers joined during the 2007 festival, compared to counties where the majority of SXSW followers had already joined before the event. The effect is persistent as Twitter adoption was higher nearly 10 years after the festival on counties with a large number of participants. 

We then use the number of new SXSW followers as an instrument (as in \cite{muller2023hashtag, gylfason2023tweets, muller2024social, fujiwara2024effect}) to address the problem of endogeneity and reverse causality. We use a Two-Stage Least Squares (2SLS) regression and we regress the suicide rate at the county level on the fitted values of the number of Twitter user on each county. We find no significant relationship between Twitter adoption and suicide rates in the United States in a 2SLS setting when we control for a large set of variables that have been found as potential drivers of suicide (unemployment, poverty, age, race, population density...).

Our paper contributes to the vast literature on the link between social media and mental health. On a large and collaborative review of the literature, \cite{haidt2023social} conclude that the evidence for a causal effect of social media is mixed. The variations in results can be attributed to several factors, including differences in methodology (e.g., cross-sectional correlational studies, longitudinal studies, and true experiments), differences in measurement (e.g., screen time, Internet use, or a focus on specific social networks), differences in the variables of interest (e.g., well-being, depression, suicide), differences in the type of social media activity (e.g., messaging friends, posting pictures, scrolling), as well as heterogeneity in usage patterns (e.g., heavy versus light users) and in the population studied (e.g., youth versus adults, men versus women). One advantage of our approach is the ability to study the long-term effects of social media development at a local level while maintaining a causal methodology. The main drawback is the external validity of our findings and the impossibility in generalizing these results based on the development of Twitter to all social networks, whose mechanisms and populations can be very different.

Empirical studies using causal designs (such as true and natural experiments) have shown a negative effect of social media on well-being. For instance, \cite{allcott2020welfare} conducted a randomized controlled trial and found that social media use (specifically Facebook) tends to reduce self-reported well-being. Similarly, \cite{braghieri2022social} identified a negative impact on well-being by analyzing the staggered introduction of Facebook across U.S. colleges. Although there is a connection between reported well-being, depression, and suicide ideation \citep{lew2019associations}, the negative effect of social media on well-being would need to be substantial to produce a noticeable impact on aggregated suicide rates. However, according to \cite{orben2019association}, the effects of digital technology use on well-being are generally small and heterogeneous, with digital use explaining at most 0.4\% of the variation in well-being. 

In this paper, we focus on suicide as our primary outcome (rather than broader measures such as mental health, anxiety or depression) primarily due to data availability. Our identification strategy requires granular data at the U.S. county level and over an extended time period (spanning before and after the introduction of Twitter). Suicide statistics are released annually by the Centers for Disease Control and Prevention (CDC) and are available at the county level from 1999 to 2020, which enable us to rigorously quantify the causal effect of Twitter's development on suicide rates.

Our paper is organized as follows. Section \ref{sec:data} presents the materials and methods. Section \ref{sec:results} presents the results of the 2SLS regressions. Section \ref{sec:discussion} discusses the results and the external validity of our approach. Section \ref{sec:conclusion} concludes.

\section{Materials \& methods}
\label{sec:data}

The South by Southwest (SXSW) festival, held annually in Austin, Texas, is widely regarded as a key event in the rapid growth and mainstream adoption of Twitter. In 2007, Twitter's presence at SXSW became a pivotal moment, as the platform was prominently showcased, leading to a significant increase in its user base and public visibility. This makes the festival an ideal setting to measure quasi-exogenous variation in Twitter’s adoption. The surge in Twitter users following the festival was largely unexpected and not directly driven by typical factors such as company-led marketing campaigns. Instead, the growth was fueled organically by the festival attendees, making it a naturally occurring event that influenced Twitter’s user base expansion. Since this spike in usage was driven by external circumstances (i.e., the SXSW event), the festival serves as a quasi-exogenous shock.

To address the endogeneity problem, we employ an instrumental variable (IV) approach, which requires two key conditions for validity: (1) the instrumental variable (presence at the SXSW festival) must be correlated with the endogenous variable (long-term social media usage); and (2) the instrumental variable must be uncorrelated with the error term in the outcome equation (suicide rates), meaning that it influences suicide rates solely through its effect on social media usage and not through any other direct or indirect channels. We downloaded the data from \cite{muller2023hashtag} available on Karsten Muller's website\footnote{https://www.karstenmueller.com/}. County-level Twitter usage in 2015 is based on a sample of 475 million geo-coded tweets collected by \cite{kinder2017archiving}. Followers of the SXSW Conference \& Festivals (@SXSW) Twitter account were web scraped : users' declared location is employed to assign each user to a county, and the date at which the user joined Twitter is used to compute the number of followers before the March 2007 festival, as well as followers that joined exactly on March 2007. We refer to Section 3 in \cite{muller2023hashtag} for a complete and detailed overview of the methodology used.

In the 2SLS regression, the first stage predicts social media usage as a function of the instrumental variable and control variables. The first-stage equation can be written as :

\begin{equation}
TwitterUser_{i,2015} = \alpha + \beta_1 \text{SXSW}_{i,March2007} + \beta_2 \text{SXSW}_{i,Pre2007} + \mathbf{X}_{i} + u_{i}
\end{equation}

where $TwitterUser_{i,2015}$ represents the logarithm of the number of Twitter users in county $i$ in 2015, $SXSW_{i,March2007}$ is the number of users who join Twitter in March 2007 (the instrumental variable) and $SXSW_{i,Pre2007}$ is the number of users who joined Twitter before March 2007 (between the creation of Twitter in March 2006 and February 2007). ${X}_{i}$ is a set of control variables including population density, county area, age distribution, gender, race distribution, poverty rate, unemployment rate and the distance to Austin, Texas (the location of the SXSW festival) (see Appendix A for a list of the controls)\footnote{Other variables, such as firearm ownership rates and drug usage patterns, could provide valuable insights due to their likely correlation with suicide rates. Unfortunately, these variables are either not available at the county level or lack sufficient longitudinal data, making their inclusion in the current model challenging.}. We also include census divisions and population deciles as fixed-effects to control for the unobservable heterogeneity.

In the second stage, the variation in the number of suicides is regressed on the predicted value of social media usage from the first stage:

\begin{equation}
\%\Delta SuicideRate_{i} = \alpha + \beta_1 \widehat{TwitterUser_{i,2015}} + \beta_2 \text{SXSW}_{i,Pre2007} + \mathbf{X}_{i} + u_{i}
\end{equation}

where $\%\Delta SuicideRate_{i}$ represents the percentage change in the suicide rate per 100,000 inhabitants. This variable is computed as the change between the mean number of suicides per county-year before the introduction of Twitter (1999 to 2005) and the mean number of suicides per county-year after the SXSW festival (from 2008 to 2020). The variable $\widehat{TwitterUser_{i,2015}}$ denotes the predicted number of Twitter users at the county level, derived from the first-stage regression of the 2SLS model. ${X}_{i}$ is a set of control variables as in Equation (1).

Data on all-cause mortality were sourced from the Underlying Cause of Death database (1999-2020) produced by the National Center for Health Statistics. Deaths attributed to suicide were identified using the 10th revision of the International Classification of Diseases (ICD-10) codes for underlying causes of death: X60–X84, Y87.0, and U03 (as in \cite{kandula2023county}). To maintain confidentiality and protect individual privacy, the absolute number of suicides is not reported by the CDC when, for a given county-year, the number of suicides falls between 1 and 9. This results in some data gaps, particularly in smaller counties where suicide counts are often missing. We present our results when we restrict the sample to counties where 10 or more suicide are observed every year over the entire period of interest (no missing observations). It represents a total of 425 counties. As a robustness check, we also present our results when we restrict the sample to counties for which we observe at least one non-missing value before 2005 and one after 2007 (1002 counties) and when we replace missing values by a random number between 1 and 9 (3058 counties).

\section{Results}
\label{sec:results}

Table 1 shows the results from the first stage regression (Equation 1) between the number of Twitter users in each county and the number of SXSW followers in March 2007. Observations are weighted by the U.S. population in 2010, and standard errors are clustered by U.S. states. We find that a 1\% increase in SXSW followers in March 2007 is associated with a 0.33\% to 0.39\% increase of the number of Twitter user in 2015. The coefficients associated with the $SXSW_{March2007}$ variable are significant in all regressions at the 1\% level while the coefficient associated with $SXSW_{Pre2007}$ is not significant when all controls are added to the model. The coefficient we estimate for the instrumental variable is smaller than the one presented in \cite{muller2023hashtag} (who find a 0.52\% increase) but the differences between the two estimates lie in the number of counties considered in the regression. We focus our first stage regression analysis on counties in which we also observe suicide data for the second stage (425 counties) whereas \cite{muller2023hashtag} use all U.S counties. However, the effect remains large and highly significant, and the F-stat in all specifications is greater than 90 which shows that our instrument is strong (see Table 2). This provides evidence on the long-term effect of the festival on the geographical adoption of Twitter.

\begin{table}[h!] \centering 
  \caption{First Stage Results - Twitter Usage (Equation 1)} 
  \label{fsresulttwt} 
\begingroup
\centering
\begin{tabular}{@{\extracolsep{5pt}} lccccc}
\\[-1.8ex]\hline 
\\[-1.8ex] & (1) & (2) & (3) & (4) & (5)  \\ 
\\[-1.8ex]
   \midrule
   \emph{Variables}\\
$SXSW_{March2007}$ & 0.3886*** & 0.3831*** & 0.3286*** & 0.336*** & 0.3268***\\
 & (0.0668) & (0.072) & (0.073) & (0.0752) & (0.07) \\
$SXSW_{Pre2007}$ & 0.1919 & 0.1978* & 0.178 & 0.1714 & 0.1937  \\
 & (0.1022) & (0.0924) & (0.0996) & (0.0976) & (0.1089) \\
    \midrule
    \emph{Controls}\\
    Geographical Controls &  & Yes & Yes & Yes & Yes  \\ 
    Demographic Controls &  &  & Yes & Yes & Yes  \\ 
    Socioeconomic Controls &  &  &  & Yes & Yes  \\ 
    Race Control &  &  &  &  & Yes \\ 
   \midrule
   \emph{Fixed-effects}\\
   Population Deciles & Yes & Yes & Yes & Yes & Yes \\  
   Census Divisions  & Yes & Yes & Yes & Yes & Yes \\  
   \midrule
   Observations & 425 & 425 & 425 & 425 & 425\\  
   Adj R$^2$ & 0.9124 & 0.922 & 0.9315 & 0.9314 & 0.9318 \\  
   \midrule \midrule
\end{tabular}
\par\endgroup
    \begin{tablenotes}
      \small \item \emph{Notes:} This table presents the results of the first-stage regression (Equation 1) at the county level. The dependent variable, $TwitterUser_{2015}$, represents the logarithm of the number of Twitter users in 2015. $SXSW_{March2007}$ is the logarithmic count of Twitter users who joined in March 2007 and followed the South by Southwest festival (SXSW), while $SXSW_{Pre2007}$ is the number of SXSW followers prior to 2007. Fixed effects for population deciles and census divisions are applied, and geographical, demographic, socioeconomic, and race controls are added incrementally. Observations are weighted by the U.S. population in 2010, and standard errors (in parentheses) are clustered by U.S. states. Significance codes: ***: 0.01, **: 0.05, *: 0.1.
       \end{tablenotes}
\end{table} 

Next, we present the results of the 2SLS regression (Equation 2) in Table 2. In the absence of controls (Model 1) or with only geographic controls (Model 2), we find that Twitter adoption is associated with a reduction in suicide rates. However, after incorporating demographic, socioeconomic, and racial controls, this relationship disappears. The coefficients for $\widehat{TwitterUser_{i,2015}}$ are not significantly different from zero in Models 3, 4, and 5. These findings indicate that there is no significant relationship between Twitter adoption and suicide rates at the county level in the United States.

\begin{table}[h!] \centering 
  \caption{Second Stage Results - Suicide Rate (Equation 2) } 
  \label{2slslnsuidiff} 
\begingroup
\centering
\begin{tabular}{@{\extracolsep{5pt}} lccccc}
\\[-1.8ex]\hline 
\\[-1.8ex] & (1) & (2) & (3) & (4) & (5)  \\ 
\\[-1.8ex]
   \midrule
$\widehat{TwitterUser_{i,2015}}$ & -0.1099* & -0.0903* & -0.052 & -0.0445 & -0.0521 \\
 & (0.0476) & (0.0378) & (0.0492) & (0.0461) & (0.0447) \\
$SXSW_{Pre2007}$ &  0.0307 & 0.0323 & 0.0305 & 0.0266 & 0.027  \\
 & (0.0234) & (0.0238) & (0.0252) & (0.0244) & (0.0259) \\
    \midrule
    \emph{Controls}\\
    Geographical Controls &  & Yes & Yes & Yes & Yes  \\ 
    Demographic Controls &  &  & Yes & Yes & Yes  \\ 
    Socioeconomic Controls &  &  &  & Yes & Yes  \\ 
    Race Control &  &  &  &  & Yes \\ 
   \midrule
   \emph{Fixed-effects}\\
   Population Deciles & Yes & Yes & Yes & Yes & Yes \\  
   Census Divisions  & Yes & Yes & Yes & Yes & Yes \\  
   \midrule
   Observations & 425 & 425 & 425 & 425 & 425 \\  
   Adj R$^2$ & 0.2296 & 0.2916 & 0.3325 & 0.355 & 0.3922 \\
   F-stat (1st stage) & 130.4 &  138.5 & 100.0 & 101.3 & 92.3\\
   \midrule \midrule
\end{tabular}

\par\endgroup
    \begin{tablenotes}
      \small \item \emph{Notes:} This table presents the results of the second-stage regression (Equation 2) at the county level. The dependent variable, $\%\Delta SuicideRate_{i}$, represents the percentage change in the suicide rate per 100,000 inhabitants between the periods 1999–2005 (pre-SXSW) and 2008–2020 (post-SXSW). The variable $\widehat{TwitterUser_{i,2015}}$ denotes the predicted number of Twitter users obtained from the first-stage regression (Equation 1), while $SXSW_{Pre2007}$ indicates the number of SXSW followers before 2007. Fixed effects for population deciles and census divisions are applied, and geographical, demographic, socioeconomic, and race controls are added incrementally. Observations are weighted by the U.S. population in 2010, and standard errors (in parentheses) are clustered by U.S. states. Significance codes: ***: 0.01, **: 0.05, *: 0.1.
    \end{tablenotes}
\end{table} 

One limitation of the results presented above is their primary focus on large counties, driven by data availability constraints. To further validate our findings, we extend the analysis by performing additional 2SLS regressions on a broader set of counties, where missing values (yearly suicide counts between 1 and 9) are imputed using two approaches: (1) replacing missing values with the average non-missing rate before and after the advent of Twitter, and (2) imputing a random number between 1 and 9. The results of these robustness checks are presented in Table 3. Across all these analyses, we consistently find no significant correlation between the predicted number of Twitter users and variations in the number of suicides.

\begin{scriptsize}
\begin{table}[h!] \centering 
  \caption{Robustness check - Missing values imputation } 
  \label{2slslnsuidiff} 
\begingroup
\centering
\begin{tabular}{@{\extracolsep{5pt}} lcc}
\\[-1.8ex]\hline 
\\[-1.8ex] & (1) & (2)   \\ 
\\[-1.8ex]
   \midrule
$\widehat{TwitterUser_{i,2015}}$ & -0.0456 & -0.0324  \\
 & (0.0275) & (0.0233)  \\
$SXSW_{Pre2007}$ &  0.0224 & 0.018   \\
 & (0.0208) & (0.0192)  \\
    \midrule
    \emph{Controls}\\
    Geographical Controls & Yes & Yes   \\ 
    Demographic Controls & Yes & Yes  \\ 
    Socioeconomic Controls & Yes & Yes  \\ 
    Race Control & Yes & Yes \\ 
   \midrule
   \emph{Fixed-effects}\\
   Population Deciles & Yes & Yes  \\  
   Census Divisions  & Yes & Yes  \\  
   \midrule
   Observations & 1002 & 3058   \\  
   Adj R$^2$ & 0.3303 & 0.1426  \\  
   \midrule \midrule
\end{tabular}

\par\endgroup
    \begin{tablenotes}
      \small \item \emph{Notes:} This table presents the results of the second-stage regression (Equation 2) at the county level. The dependent variable, $\%\Delta SuicideRate_{i}$, represents the percentage change in the suicide rate per 100,000 inhabitants between the periods 1999–2005 (pre-SXSW) and 2008–2020 (post-SXSW). The variable $\widehat{TwitterUser_{i,2015}}$ denotes the predicted number of Twitter users obtained from the first-stage regression (Equation 1), while $SXSW_{Pre2007}$ indicates the number of SXSW followers before 2007. In Model [1], missing values are imputed using the mean number of suicides observed before and after the period of interest. In Model [2], missing values are imputed with a random number between 1 and 9. Fixed effects for population deciles and census divisions are applied, and geographical, demographic, socioeconomic, and race controls are added incrementally. Observations are weighted by the U.S. population in 2010, and standard errors (in parentheses) are clustered by U.S. states. Significance codes: ***: 0.01, **: 0.05, *: 0.1.
    \end{tablenotes}
\end{table} 
\end{scriptsize}

\section{Discussion}
\label{sec:discussion}

Our findings suggest that Twitter adoption neither increases nor decreases suicide rates in the United States. To the best of our knowledge, this represents the first empirical evidence demonstrating the absence of an effect from the platform's expansion on suicide rates, utilizing a causal identification strategy based on the platform's spatial development. This result remains robust even after including extensive control variables and employing various methodologies for imputing missing values.

We would like to emphasize that our findings should be interpreted with caution and not over-generalized. First, Twitter's format and interaction patterns differ significantly from other platforms such as Facebook, Instagram, or TikTok. Twitter primarily emphasizes short, text-based posts and (mostly) public conversations, whereas other platforms focus more on images or videos and offer more private interactions. These distinctive features may result in differing impacts on user behavior and emotional well-being, limiting the external validity of our findings to other social media platforms. Second, due to data limitations, we were unable to analyze the impact on specific subpopulations, particularly teenagers, who are often identified as being more vulnerable to the effects of heavy social media usage. As noted earlier, the CDC does not publish suicide counts for counties where the number of suicides falls between 1 and 9. Since the annual number of teenage suicides per county is below this threshold in almost all cases, we could not perform a subgroup analysis. Third, our analysis does not account for the rapid evolution of social media platforms and their algorithms over time, which can significantly influence user experiences and societal impacts. The structure and influence of Twitter during the period we studied (before 2020) may differ significantly from its current state, particularly after its acquisition by Elon Musk in 2022 as disinformation and harmful content on Twitter/X have increased in recent years \citep{hickey2023auditing}. Fourth, our approach does not consider the quality or tone of interactions on Twitter, which can vary widely and influence its psychological effects. Positive outcomes, such as exposure to supportive communities or mental health resources, may mitigate negative effects, while experiences with cyberbullying, misinformation, or distressing content could exacerbate them. The null effect we identify may be the result of these opposing forces balancing out. This does not negate the possibility that some individuals were significantly negatively impacted by their social media use. However, at an aggregate level, these negative impacts may be offset by the benefits experienced by others.

\section{Conclusion}
\label{sec:conclusion} 

Using Twitter's quasi-exogenous geographic expansion following the 2007 SXSW festival and employing an instrumental variable approach, we find no significant impact of Twitter adoption on county-level suicide rates, after accounting for various demographic, geographic and socioeconomic factors. This absence of effect carries significant implications for public policy, emphasizing the need for a balanced approach that carefully considers the diverse impacts of social media. It is crucial to recognize that not all social media platforms are the same. While it cannot be asserted that social media platforms uniformly exert no effects on mental health outcomes, it is equally important to demonstrate that some platforms—such as Twitter, at least in the period before its 2022 acquisition by Elon Musk—exhibit no measurable impact. Given the limitations of this study’s external validity, future research should build on these findings by exploring other platforms and populations to better capture the heterogeneous effects of social media. Additionally, examining the evolving nature of platform algorithms and user interactions over time will be critical to understanding the broader implications of social media on mental health outcomes.

\clearpage

\pagebreak
\singlespacing
\bibliographystyle{apalike}
\bibliography{article}

\pagebreak	

\onehalfspacing

\section*{Appendix A}

This appendix provides details on the control variables included in the 2SLS regression, organized by category: Fixed Effects, Geographic Controls, Demographic Controls, Socioeconomic Controls, and Race Controls. All control variables are constructed prior to the creation of Twitter. \\

\textbf{Fixed Effects}: We incorporate population deciles, as outlined in \cite{muller2023hashtag}, based on population data from the U.S. Census Bureau. Additionally, we include the nine U.S. Census Divisions: New England, Middle Atlantic, East North Central, West North Central, South Atlantic, East South Central, West South Central, Mountain, and Pacific. \\

\textbf{Geographic Controls}: The area of each county is sourced from the U.S. Census Bureau database. Population density is calculated by dividing the population of each county by its area. We also compute the distance to Austin, Texas, using the Haversine formula based on the latitude and longitude of each county centroid. \\

\textbf{Demographic Controls}: We account for the age distribution using U.S. Census Bureau data and include the following groups in the model: 18–24 years, 25–44 years, 45–64 years, and 65+ years. The percentage of women in the population is calculated from the Population Estimates by Age, Sex, and Race dataset. \\

\textbf{Socioeconomic Controls}: The county-level unemployment rate is obtained from the U.S. Bureau of Labor Statistics. County poverty rates are sourced from the U.S. Census Bureau. \\

\textbf{Race Controls}: We calculate the percentage of the population identified as White and Black using data from the 2000 Census (Population Estimates by Age, Sex, and Race), \\

\end{document}